\documentclass[10pt]{article}
\usepackage[dvips]{graphicx}
\usepackage{amsmath,amsfonts,amssymb,latexsym,epsfig}
\usepackage{mathrsfs}
\usepackage{verbatim}
\usepackage{latexsym}
\usepackage{amsthm}
\usepackage{amssymb}
\usepackage{graphics}
\usepackage{amsbsy}
\usepackage{enumerate}
\usepackage{times}
\numberwithin{equation}{section}
\graphicspath{{../../brown_mot/Freidlin_Wentzell/}}

\newcommand{\E}{{\mathbb E}}

\newcommand{\LL}{{\mathcal L}}

\newcommand{\T}{[-\pi, \pi]}
\newcommand{\R}{{\mathbb R  }}

\newcommand{\pdt}[1]{\frac{\partial #1}{\partial t}}

\newcommand{\pddd}[3]{\frac{\partial^2 #1}{\partial {#2} \partial{#3}}}

\newcommand{\brk}[1]{\left( #1 \right)}

\newcommand{\cL}{\mathcal L}

\newcommand{\eps}{\epsilon}

\newcommand{\Deff}{D_{eff}}

\newcommand{\fnk}[2]{\ensuremath{\phi _{#1\,#2}}}
\begin{document}
\title{DIFFUSIVE TRANSPORT IN PERIODIC POTENTIALS: UNDERDAMPED DYNAMICS}
\author{ G.A. Pavliotis \\
        Department of Mathematics\\
    Imperial College London \\
        London SW7 2AZ, UK \\ and \\
    T. Vogiannou \\
    Department of Physics \\
    Aristotle University of Thessaloniki \\
    54124 Thessaloniki, Greece
                }
\maketitle
\begin{abstract}
In this paper we present a systematic and rigorous method for calculating the diffusion
tensor for a Brownian particle moving in a periodic potential which is valid in arbitrary
dimensions and for all values of the dissipation. We use this method to obtain an
explicit formula for the diffusion coefficient in one dimension which is valid in the
underdamped limit, and we also obtain higher order corrections to the Lifson-Jackson
formula for the diffusion coefficient in the overdamped limit. A numerical method for
calculating the diffusion coefficient is also developed and is shown to perform extremely
well for all values of the dissipation.

\end{abstract}
%
%%%%%%%%%%%%%%%%%%%%%%%%%%%%%%%%%%%%%%%%%%%%%%%%%%%%%%%%%%%%%%%%%%%%%%%%%%%%%%
%
%                                      INTRODUCTION
%
%%%%%%%%%%%%%%%%%%%%%%%%%%%%%%%%%%%%%%%%%%%%%%%%%%%%%%%%%%%%%%%%%%%%%%%%%%%%%%
%
\section{Introduction}
\label{sec:intro}
Brownian motion in periodic and random potentials has been a very active area of research
for many decades. Apart from the well established applications to electronics
\cite{straton63, straton67} and to solid state physics such as superionic conductors, the
Josephson tunneling junction \cite{BarPater82} and surface diffusion~\cite{gomer90},  new
and exciting applications to physics (self-assembled molecular film growth, catalysis,
surface-bound nanostructures) and to biology (stochastic modeling of molecular and
Brownian motors~\cite{reimann}) keep the subject of Brownian motion at the forefront of
current research, both theoretical and experimental.

Despite the fact that Brownian motion in periodic potentials has been studied
extensively~\cite[Ch. 11]{Ris84}, \cite{coffey04} and many analytical and numerical
results have been obtained, there are still many open questions, in particular in the
underdamped, multidimensional case. The main purpose of this paper is to develop a
general method for calculating the diffusion tensor $D$ of the Brownian particle in
arbitrary dimensions, and to then use this method for setting up an efficient numerical
method for computing $D$. Furthermore, we will show that our method for calculating $D$
will enable us to study various asymptotic limits of physical interest in a systematic
and rigorous fashion.

The dynamics of a Brownian particle moving in a periodic potential is governed by the
Langevin equation
\begin{equation}\label{e:langevin}
\ddot{x} = - \nabla V(x) - \gamma \dot{x} + \xi,
\end{equation}
where $x(t)$ denotes the particle position, $V(x)$ is a smooth periodic potential,
$\gamma$ denotes the friction coefficient and $\xi(t)$ is a white noise Gaussian
 process with correlation function
$$
\langle \xi_i(t) \xi_j(s) \rangle = 2 \gamma k_B T  \delta_{ij} \delta(t-s), \quad
i,j=1,\dots,d.
$$
where $k_B$ is Boltzmann's constant and $T$ is the absolute temperature, in accordance
with the fluctuation-dissipation theorem. We use the notation $\langle \cdot \rangle$ to
denote ensemble average. We will also write $\beta = (k_B T)^{-1}$ and $\xi(t) = \sqrt{2
\gamma \beta^{-1}} \dot{W}$, where $W(t)$ is a standard Brownian motion in $\R^d$.
Throughout this work we will assume that the diffusing particle is of unit mass, $m=1$.

The Langevin equation~\eqref{e:langevin} has been studied extensively as a theoretical
model for the diffusion of adsorbates on crystal surfaces~\cite{gomer90,pollak05}. In
this setting, $q(t)$ represents the position of the diffusing particle, $V(q)$ the
substrate potential and the friction and noise terms represent the interaction of the
diffusing particle with the phonon heat bath~\cite{pollak05}

It is well known that at low temperatures (which is usually the regime of physical
interest), the diffusing particle performs a hopping motion (random walk) between the
local minima of the potential. This hopping motion is characterized by the mean square
jump length $\langle \ell^2 \rangle$ and the hopping rate $\kappa$ (or, equivalently, the
mean escape time $\tau$ with $\kappa = \frac{1}{2 \tau}$). These two quantities are
related to the diffusion coefficient through the formula (in one dimension)
\begin{equation}\label{e:deff_jump}
D = \langle \ell^2 \rangle \kappa.
\end{equation}
Knowledge of two of the three quantities $(D, \, \langle \ell^2 \rangle, \, \kappa )$ is
sufficient for the calculation of the third. Of course, the diffusion tensor can also be
defined either in terms of the ensemble average of the second moment, i.e.
\begin{equation}\label{e:deff_langr}
D = \lim_{t \rightarrow \infty} \frac{1}{2 t} \langle \big(q(t) -q(0) \big) \otimes (q(t)
- q(0)) \rangle,
\end{equation}
or in terms of the time integral of the velocity autocorrelation function, i.e. through
the Green-Kubo formula
\begin{equation}\label{e:green_kubo}
D = \int_0^{\infty} \langle p(t) \otimes p(0) \rangle \, dt
\end{equation}
with $p(t) = \dot{q}(t)$. In the above formulas $\otimes$ stands for the tensor product
between two vectors in $\R^d$. We remark that, although formulas~\eqref{e:deff_langr}
and~\eqref{e:green_kubo} are valid in arbitrary dimensions, it is not clear how to
interpret formula~\eqref{e:deff_jump} in dimensions higher than $1$.

Of particular interest is the dependence of the diffusion coefficient $D$  as well as the
jump rate $\kappa$  and (mean square) jump length $\langle\ell^2 \rangle$  on the
friction coefficient. In particular, it is well known that in the underdamped regime the
particle diffusion is dominated by the occurrence of long
jumps~\cite{Braun_Ferrando2002,Chen_al1996, GuVeMiPo03}. Apart from theoretical
investigations and numerical simulations based on the Langevin
dynamics~\eqref{e:langevin}, the occurrence of long jumps in the underdamped regime is
also verified by means of molecular dynamics simulations of a model for
CO/Ni$(111)$~\cite{DobDor92} and is also, by now, a well established experimental
result~\cite{SchLindRosLaegStenBes02, pollak05}. Indeed, in the case of weak
adsorbate-substrate interaction i.e. in the case of weak coupling between the diffusing
particle and the heat bath which corresponds to the underdamped dynamics regime, the
diffusion mechanism is controlled by long jumps, spanning multiple lattice
spacings~\cite{SchLindRosLaegStenBes02, pollak05}. The rigorous and systematic study of
the diffusion process in the weak dissipation regime is still a major challenge for
theoreticians, in particular in dimensions higher than $1$.

The problem of diffusion in periodic potentials is well studied in one
dimension~\cite{Ris84,Braun_Ferrando2002, pollak05} or for separable potentials in two
and three dimensions~\cite{sancho_al04b}. Kramers' theory~\cite{Kramers40} applied to
Brownian motion in a periodic potential or the mean first passage time
method~\cite{Schuss81} enables us to calculate the rate of escape from a local minimum of
the potential (hopping rate). In the overdamped limit this is sufficient to calculate the
diffusion coefficient, since only single jumps occur and consequently $\langle \ell^2
\rangle = L^2$ where $L$ is the period of the potential. This leads to the well known
Lifson-Jackson formula for the diffusion coefficient~\cite{lifson_jackson62}\cite[Ch.
13]{PavlSt08}
\begin{equation}\label{e:lifson_jackson}
D  = \frac{D_0 L^2}{\int_0^L e^{\beta V(q)} \, dq \int_0^L e^{-\beta V(q)}},
\end{equation}
where $\beta^{-1} = k_B T$ and $D_0$ denotes the diffusion coefficient of the free
particle
$$
D_0 = \frac{k_B T}{\gamma}.
$$
Kramers' formula for the rate of escape enables us to obtain a formula for the diffusion
coefficient which is valid in the moderate-to-strong friction regime, for small
temperatures $\beta \gg 1$ ~\cite{GuVeMiPo03,Kozlov89}:
\begin{equation}\label{e:kozlov}
D = \frac{\omega_0}{2 \pi} \left(\sqrt{1 + \frac{\gamma^2}{4 \omega_b^2}} -
\frac{\gamma}{2 \omega_b} \right) e^{- \beta E_b},
\end{equation}
where $E_b = V(q_{MAX}) - V(q_{MIN}), \; \omega_0^2 = V''(q_{MIN}), \; \omega_b^2 =
|V''(q_{MAX})|$. In the overdamped, $\gamma/\omega_b \gg 1$, small temperature $\beta \gg
1$, limit this formula reduces to the small temperature asymptotics of
equation~\eqref{e:lifson_jackson}:
\begin{equation}\label{e:lif_jack_beta}
D = \frac{\omega_0 \omega_b L^2}{2 \pi \gamma} e^{- \beta E_b}.
\end{equation}
The calculation of the diffusion coefficient in the underdamped limit requires two
calculations, that of the hopping rate and that of the mean squared jump length $\langle
\ell^2 \rangle$. Such a calculation was presented in~\cite{sancho_al04a, sancho_al04b}
for the case of a cosine potential. For this potential, a formula for the diffusion
coefficient which is valid in the regime $\gamma \ll 1$ was also obtained by Risken and
presented in his monograph~\cite{Ris84}:
\begin{equation}\label{e:risken}
D = \frac{1}{\gamma} \frac{\pi}{2 \beta} e^{-2 \beta}.
\end{equation}
In contrast to the one dimensional problem, a similar theory in higher dimensions is
still lacking, except for the overdamped limit. In this limit approximate analytical
results for certain two-dimensional potentials have been derived in the
literature~\cite{ferrando_all92}. Furthermore,  it is possible to prove, using rigorous
mathematical analysis, that the diffusion tensor scales like $\frac{1}{\gamma}$ when
$\gamma \gg 1$~\cite{HP07}.

On the contrary, it is still not clear what the scaling of the (trace of the) diffusion
tensor with the friction constant is in the underdamped, multidimensional case. Numerical
experiments~\cite{Braun_Ferrando2002} suggest that this scaling depends crucially on the
detailed properties of the periodic  potential; however, a rigorous and systematic theory
for explaining the dependence of the diffusion coefficient on the strength of the
dissipation in arbitrary dimensions is still lacking.

The situation becomes even more unclear when an external driving force (either constant
or periodic in time) is present. Brownian motion in tilted periodic potentials has only
been studied in one dimension~\cite{Ris84} and explicit formulas are only valid in the
overdamped limit~\cite{reimann_al01, lindner_al01, pavl05}. Furthermore, numerical
experiments~\cite{ZhangBao03} seem to indicate that stochastic resonance in periodic
potentials in only possible in dimensions higher than one.

In view of the ubiquity of diffusive motion in periodic potentials in applications,
 it seems to be important to develop the multidimensional theory in a systematic
 and rigorous way, taking into account external driving forces. This paper
 is a contribution towards this goal, and it is a part of our research program
 on the study of Brownian motion in periodic and random potentials~\cite{PavlSt06,
 HP07, pavl05}.

All studies on the problem of Brownian motion in a periodic potential that have been
reported in the physics literature rely heavily on the analysis of the Fokker-Planck
(Kramers-Chandrashekhar) equation which governs the evolution of the transition
probability density for the Brownian particle. For example, the continued fraction
expansion method is used in order to solve the Fokker-Planck equation~\cite{coffey04,
Ris84} in a semi-analytic fashion and to calculate quantities such as the mobility, the
intermediate structure function and the dynamic structure function~\cite{DPS77}. Or,
Kramers' theory is being used, which again relies on the study of the Fokker-Planck
equation.

On the other hand, many tools from stochastic analysis~\cite{RevYor99}, the theory of
limit theorems for Markov processes~\cite{EthKur86} and the emerging field of multiscale
analysis~\cite{PavlSt08} are appropriate for the study of this problem and, yet, they
have received very little attention in the physics community. The purpose of this paper
is to use multiscale methods such as homogenization theory and singular perturbation
theory in order to offer a new insight into the problem of Brownian motion in periodic
potentials. In particular, we derive in a rigorous and systematic way a formula for the
diffusion tensor $D$ of a Brownian particle moving in a periodic potential in arbitrary
dimensions and then we use in order to develop an efficient numerical method for
calculating $D$. This numerical method is related to the continued fraction expansion
method, but is easier to implement and to analyze. As a byproduct of our analysis, we
derive rigorously a formula for the diffusion coefficient which is valid in the weak
friction limit; furthermore, we also calculate higher order corrections to the large
$\gamma$ asymptotics of the diffusion coefficient.

The rest of the paper is organized as follows. In Section~\ref{sec:multiscale} we present
the multiscale analysis and we derive a formula for the diffusion coefficient. We also
show the equivalence between our formula and the Green-Kubo formula~\eqref{e:green_kubo}.
In Section~\eqref{sec:asymptotics} we derive formulas which are valid in the $\gamma
\rightarrow 0$ and $\gamma \rightarrow + \infty$ limits. In Section~\ref{sec:numerical}
we develop the numerical method and we compare the numerical results obtained using our
method with results obtained from Monte Carlo simulations, from approximate analytical
formulas and from the numerical implementation of formula~\eqref{e:deff_jump}.
Section~\ref{sec:conclusions} is reserved for conclusions.
%%%%%%%%%%%%%%%%%%%%%%%%%%%%%%%%%%%%%%%%%%%%%%%%%%%%%%%%%%%%%%%%%%%%%%%%%%%%%%
%
\section{Multiscale Analysis}
\label{sec:multiscale}
In this section we use multiscale analysis~\cite{PavlSt08} to derive a formula for the
diffusion tensor of a Brownian particle moving in a periodic potential in arbitrary
dimensions. We then show the equivalence between this formula and the Green-Kubo formula
for the diffusion tensor.
%%%%%%%%%%%%%%%%%%%%%%%%%%%%%%%%%%%%%%%%%%%%%%%%%%%%%%%%%%%%%%
\subsection{Derivation of Formula for the Diffusion Tensor}
We start by rescaling the Langevin equation~\eqref{e:langevin}
\begin{equation}\label{e:langevin_2}
\ddot{x} = F(x) - \gamma \dot{x} + \sqrt{2 \gamma \beta^{-1}} \dot{W},
\end{equation}
where we have set $F(x) = - \nabla V(x)$. We will assume that the potential is periodic
with period $2 \pi$ in every direction.
 Since we expect that at
sufficiently long length and time scales the particle performs a purely diffusive motion,
we perform a diffusive rescaling to the equations of motion~\eqref{e:langevin}: $t
\rightarrow t/\epsilon^2$, $x \rightarrow \frac{x}{\epsilon}$. Using the fact that
$\dot{W}(c \, t) = \frac{1}{\sqrt{c}} \dot{W}(t)$ in law we obtain:
\begin{equation}
 \epsilon^2 \ddot{x} = \frac{1}{\epsilon} F \left( \frac{x}{\epsilon}\right) -
 \gamma \dot{x} + \sqrt{2 \gamma \beta^{-1}} \dot{W},
\nonumber
\end{equation}
 Introducing $p= \epsilon \dot{x}$ and
$q=x/\epsilon$ we write this equation as a first order system:
\begin{equation}
\begin{array}{ccc}
\dot{x} & = &\frac{1}{ \epsilon} p,\\
\dot{p} & = &\frac{1}{\epsilon^2} F(q) - \frac{1}{ \epsilon^2} \gamma  p +
\frac{1}{\epsilon^2}\gamma \beta^{-1} \dot{W},\\
\dot{q} & = &\frac{1}{\epsilon^2} p,
\end{array}
\label{eqn:langevin_rescaled}
\end{equation}
with the understanding that $q \in [-\pi, \pi]^d$ and $x, \, p \in {\mathbb R}^d.$ Our
goal now is to eliminate the fast variables $p, \, q$ and to obtain an equation for the
slow variable $x$. We shall accomplish this by studying the corresponding backward
Kolmogorov equation using singular perturbation theory for partial differential
equations.

Let
$$u^{\eps}(p,q,x,t) = \E f\big(p(t), q(t), x(t)|p(0)=p, q(0)= q, x(0)=x \big),$$
where $\E$ denotes the expectation with respect to the Brownian motion $W(t)$ in the
Langevin equation and $f$ is a smooth function.\footnote{In other words, we have that
$$
u^{\eps}(p,q,x,t) =  \int f(x,v,t;p,q) \rho(x,v,t; p,q) \mu(p,q) \, dp dq dx dv,
$$
where $\rho(x,v,t;p,q)$ is the solution of the Fokker-Planck equation and $\mu(p,q)$ is
the initial distribution.

} The evolution of the function $u^{\eps}(p,q,x,t)$ is governed by the backward
Kolmogorov equation associated to equations~\eqref{eqn:langevin_rescaled}
is~\cite{PavlSt08}\footnote{it is more customary in the physics literature to use the
forward Kolmogorov equation, i.e. the Fokker-Planck equation. However, for the
calculation presented below, it  is more convenient to use the backward as opposed to the
forward Kolmogorov equation. The two formulations are equivalent. See \cite[Ch.
6]{PavlSt06} for details. }
\begin{eqnarray}
\frac{\partial u^{\epsilon}}{\partial t} & = & \frac{1}{\epsilon} p \cdot \nabla_x
u^{\epsilon} + \frac{1}{\epsilon^2} \Big( -\nabla_q V(q) \cdot \nabla_p + p \cdot
\nabla_q + \gamma \big(- p \cdot \nabla_p + \beta^{-1} \Delta_p \big) \Big) u^{\epsilon}.
    \nonumber \\ & := &
        \left( \frac{1}{\epsilon^2} \mathcal{L}_0 +
        \frac{1}{\epsilon} \mathcal{L}_1  \right)  u^{\epsilon},
\label{eq:backw_kolmog_2}
\end{eqnarray}
where:
\begin{eqnarray*}
\mathcal{L}_0 & = &  -\nabla_q V(q) \cdot \nabla_p + p \cdot \nabla_q + \gamma \big(- p
\cdot \nabla_p + \beta^{-1}
\Delta_p \big),\\
\mathcal{L}_1 & = &  p \cdot \nabla_x
\end{eqnarray*}
The invariant distribution of the fast process $\big\{ q(t), \, p(t) \big\}$ in $\T^d
\times \R^d$ is the Maxwell-Boltzmann distribution
$$
\rho_{\beta}(q, p) = Z^{-1} e^{-\beta H(q,p)}, \quad Z = \int_{\T^d \times \R^d}
e^{-\beta H(q,p)} \, dq dp,
$$
where $H(q, p) = \frac{1}{2}|p|^2 + V(q)$. Indeed, we can readily check that
$$
\cL^*_0 \rho_{\beta}(q,p) = 0,
$$
where $\cL^*_0$ denotes the Fokker-Planck operator which is the $L^2$-adjoint of the
generator of the process $\LL_0$:
$$
\cL_0^*f \cdot = \nabla_q V(q) \cdot \nabla_p f - p \cdot \nabla_q f+ \gamma \big(
\nabla_p \cdot (p f ) + \beta^{-1} \Delta_p f \big).
$$
The null space of the generator $\LL_0$ consists of constants in $q, \, p$. Moreover, the
equation
\begin{equation}\label{e:poisson}
-\mathcal{L}_0 f = g,
\end{equation}
has a unique (up to constants) solution if and only if
\begin{equation}
\label{eq:int0} \langle g \rangle_{\beta} := \int_{\T^d \times \R^d} g(q,p)
\rho_{\beta}(q, p) \, dq dp = 0.
\end{equation}
Equation~\eqref{e:poisson} is equipped with periodic boundary conditions with respect to
$z$ and is such that
\begin{equation}\label{e:l2_fin}
\int_{\T^d \times \R^d} |f|^2 \mu_{\beta} \, dq dp < \infty.
\end{equation}
These two conditions are sufficient to ensure existence and uniqueness of solutions (up
to constants) of equation~\eqref{e:poisson} \cite{HP07, HairPavl04, papan_varadhan}.

We assume that the following ansatz for the solution $u^{\epsilon}$ holds:
\begin{equation}
u^{\epsilon} = u_0 + \epsilon u_1 + \epsilon^2 u_2 + \dots \label{eq:expansion}
\end{equation}
with $u_i = u_i(p,q,x, t), \, i=1,2, \dots$ being $2 \pi$ periodic in $q$ and satisfying
condition~\eqref{e:l2_fin}. We substitute (\ref{eq:expansion}) into
(\ref{eq:backw_kolmog_2}) and equate equal powers in $\eps$ to obtain the following
sequence of equations:
\begin{subequations}
\begin{eqnarray}
\mathcal{L}_0 \, u_0 & = &0,\\
\mathcal{L}_0 \, u_1 & = &- \mathcal{L}_1 \, u_0,\\
\mathcal{L}_0 \, u_2 & = &- \mathcal{L}_1 \, u_1 + \frac{\partial u_0}{\partial t}.
\end{eqnarray}
\label{eq:oep}
\end{subequations}
From the first equation in (\ref{eq:oep}) we deduce that $u_0 = u_0(x,t)$, since the null
space of $\LL_0$ consists of functions which are constants in $p$ and $q$. Now the second
equation in (\ref{eq:oep}) becomes:
\begin{equation*}
\mathcal{L}_0 u_1 = - p \cdot \nabla_x u_0.
\end{equation*}
Since $\langle p \rangle =0$, the right hand side of the above equation is mean-zero with
respect to the Maxwell-Boltzmann distribution. Hence, the above equation is well-posed.
We solve it using separation of variables:
\begin{equation*}
u_1 = \Phi(p,q) \cdot \nabla_x u_0
\end{equation*}
with
\begin{equation}
-\mathcal{L}_0 \Phi =p. \label{eqn:cell}
\end{equation}
This Poisson equation is posed on $\T^d \times \R^d$. The solution is periodic in $q$ and
satisfies condition~\eqref{e:l2_fin}. Now we proceed with the third equation in
\eqref{eq:oep}. We apply the solvability condition to obtain:
\begin{eqnarray*}
\frac{\partial u_0}{\partial t} &  =  &  \int_{\T^d \times \R^d} \mathcal{L}_1 u_1
\rho_{\beta}(p,q) \, dp dq \\ & = & \sum_{i,j=1}^d \left( \int_{\T^d \times \R^d} p_i
\Phi_j \rho_{\beta}(p,q) \, dp dq \right)  \pddd{u_0}{x_i}{x_j}.
\end{eqnarray*}
This is the Backward Kolmogorov equation which governs the dynamics on large scales. We
write it in the form
\begin{equation}
\frac{\partial u_0}{\partial t} =  \sum_{i,j=1}^d D_{ij} \pddd{u_0}{x_i}{x_j}
\label{eqn:homog_lang}
\end{equation}
where the effective diffusion tensor is
\begin{equation}
D_{ij} = \int_{\T^d \times \R^d} p_i \Phi_j \rho_{\beta}(p,q) \, dp dq, \quad i,j=1,
\dots d. \label{eqn:eff_diff}
\end{equation}
The calculation of the effective diffusion tensor requires the solution of the boundary
value problem \eqref{eqn:cell} and the calculation of the integral
in~\eqref{eqn:eff_diff}. The limiting backward Kolmogorov equation is well posed since
the diffusion tensor is nonnegative. Indeed, let $\xi$ be a unit vector in $\R^d$. We
calculate (we use the notation $\Phi_{\xi} = \Phi \cdot \xi$ and $\langle \cdot, \cdot
\rangle$ for the Euclidean inner product.)
\begin{eqnarray}
\langle \xi , D \xi \rangle &=& \int (p \cdot \xi) (\Phi_{\xi} ) \mu_{\beta} \, dp dq =
\int \big(- \cL_0 \Phi_{\xi} \big) \Phi_{\xi}  \mu_{\beta} \, dp dq \nonumber \\ & = &
\gamma \beta^{-1} \int \big|\nabla_p \Phi_{\xi} \big|^2 \mu_{\beta} \, dp dq \geq 0,
\label{e:pos_def}
\end{eqnarray}
where an integration by parts was used.

Thus, from the multiscale analysis we conclude that at large lenght/time scales the
particle which diffuses in a periodic potential performs and effective Brownian motion
with a nonnegative diffusion tensor which is given by formula~\eqref{eqn:eff_diff}.

We mention in passing that the analysis presented above can also be applied to the
problem of Brownian motion in a tilted periodic potential. The Langevin equation becomes
\begin{equation}\label{e:langevin_intro}
\ddot{x}(t) = - \nabla V(x(t)) +F - \gamma \dot{x}(t) + \sqrt{2 \gamma \beta^{-1}}
\dot{W}(t),
\end{equation}
where $V(x)$ is periodic with period $2 \pi$ and $F$ is a constant force field. The
formulas for the effective drift and the effective diffusion tensor are
\begin{equation}\label{e:veff_deff}
V = \int_{\R^d \times \T^d} p \rho(q,p) \, dq dp, \quad D = \int_{\R^d \times \T^d} (p -
V) \otimes  \phi \rho(p,q) \, dp dq,
\end{equation}
where
\begin{subequations}\label{e:cell_fp}
\begin{equation}\label{e:cell}
- \cL \phi = p-V,
\end{equation}
\begin{equation}\label{e:fp}
 \cL^* \rho = 0, \quad \int_{\R^d \times \T^d} \rho(p,q) \, dp dq = 1.
 \end{equation}
\end{subequations}
with
\begin{equation}\label{e:generator}
\cL = p \cdot \nabla_q + (-\nabla_q V + F) \cdot \nabla_p + \gamma \big( - p \cdot
\nabla_p + \beta^{-1} \Delta_p \big).
\end{equation}
We have used $\otimes$ to denote the tensor product between two vectors; $\cL^*$ denotes
the $L^2$-adjoint of the operator $\cL$, i.e. the Fokker-Planck operator.
Equations~\eqref{e:cell_fp} are equipped with periodic boundary conditions in $q$. The
solution of the Poisson equation~\eqref{e:cell_fp} is also taken to be square integrable
with respect to the invariant density $\rho(q,p)$:
$$
\int_{\R^d \times \T^d} |\phi(q,p)|^2 \rho(p,q) \, dp dq <  +\infty.
$$
The diffusion tensor is nonnegative definite. A calculation similar to the one used to
derive~\eqref{e:pos_def} shows the positive definiteness of the diffusion tensor:
\begin{eqnarray}
\langle \xi , D \xi \rangle = \gamma \beta^{-1} \int \big|\nabla_p \Phi_{\xi} \big|^2
\rho(p,q) \, dp dq \geq 0,
\end{eqnarray}
for every vector $\xi$ in $\R^d$. The study of diffusion in a tilted periodic potential,
in the underdamped regime and in high dimensions, based on the above formulas for $V$ and
$D$, will be the subject of a separate publication.
%%%%%%%%%%%%%%%%%%%%%%%%%%%%%%%%%%%%%%%%%%%%%%%%%%%%%%%%%%%%%%%%%%%%%%%%%%%%%%%%
\subsection{Equivalence With the Green-Kubo Formula}
Let us now show that the formula for the diffusion tensor obtained in the previous
section, equation~\eqref{eqn:eff_diff}, is equivalent to the Green-Kubo
formula~\eqref{e:green_kubo}. To simplify the notation we will prove the equivalence of
the two formulas in one dimension. The generalization to arbitrary dimensions is
immediate. Let $(x(t;q,p), \, v(t;q,p) )$ with $v = \dot{x}$ and initial conditions
$x(0;q,p) = q, \, v(0;q,p) = p$ be the solution of the Langevin equation
$$
\ddot{x} = - \partial_x V - \gamma \dot{x} + \xi
$$
where $\xi(t)$ stands for Gaussian white noise in one dimension with correlation function
$$
\langle \xi(t) \xi(s) \rangle = 2 \gamma k_B T \delta(t-s).
$$
We assume that the $(x, \, v)$ process is stationary,  i.e. that the initial conditions
are distributed according to the Maxwell-Boltzmann distribution
$$
\rho_{\beta}(q,p) = Z^{-1}e^{-  \beta H(p,q)}.
$$
The velocity autocorrelation function is~\cite[eq. 2.10]{DPS77}
\begin{equation}\label{e:correl}
\langle v(t;q,p) v(0;q,p) \rangle = \int v  \, p \rho(x,v,t; p,q) \rho_{\beta}(p,q) \, dp
dq dx dv,
\end{equation}
and $\rho(x,v,t;p,q)$ is the solution of the Fokker-Planck equation
$$
\pdt{\rho} = \cL^* \rho, \quad \rho(x,v,0;p,q) = \delta(x-q) \delta(v-p),
$$
where
$$
\cL^* \rho = - v \partial_x \rho + \partial_x V(x) \partial_v \rho + \gamma \big(
\partial (v \rho) + \beta^{-1} \partial_v^2 \rho \big).
$$
We rewrite~\eqref{e:correl} in the form
\begin{eqnarray}
\langle v(t;q,p) v(0;q,p) \rangle & = & \int \int \left( \int \int v \rho(x,v,t; p,q) \,
dv dx \right) p \rho_{\beta}(p,q) \, dp dq \nonumber
\\ & =: & \int \int \overline{v}(t; p,q) p \rho_{\beta}(p,q) \, dp dq. \label{e:corel_2}
\end{eqnarray}
The function $\overline{v}(t)$ satisfies the backward Kolmogorov equation which governs
the evolution of observables~\cite[Ch. 6]{PavlSt08}
\begin{equation}\label{e:back_kolm_v}
\pdt{\overline{v}} = \cL \overline{v}, \quad v(0; p,q) = p.
\end{equation}
We can write, formally, the solution of~\eqref{e:back_kolm_v} as
\begin{equation}\label{e:semigroup}
\overline{v} = e^{\cL t} p.
\end{equation}
We combine now equations~\eqref{e:corel_2} and~\eqref{e:semigroup} to obtain the
following formula for the velocity autocorrelation function
\begin{equation}\label{e:vel_corel_2}
\langle  v(t;q,p) v(0;q,p)  \rangle = \int \int p \big( e^{\cL t} p \big)
\rho_{\beta}(p,q) \, dp dq.
\end{equation}
We substitute this into the Green-Kubo formula to obtain
\begin{eqnarray*}
D & = & \int_0^{\infty} \langle v(t;q,p) v(0;q,p) \rangle \, dt \\
& = & \int \left( \int_0^{\infty} e^{\cL t} \, dt \, p \right) p \rho_{\beta} \, dp dq \\
& = & \int \Big(- \cL^{-1} p \Big) p \rho_{\beta} \, dp dq \\
& = & \int_{-\infty}^{\infty} \int_{-\pi}^{\pi} \phi p \rho_{\beta} \, dp dq,
\end{eqnarray*}
where $\phi$ is the solution of the Poisson equation~\eqref{eqn:cell}. In the above
derivation we have used the formula $-\cL^{-1} = \int_0^{\infty} e^{\cL t} \, dt$, whose
proof can be found in~\cite[Ch. 11]{PavlSt08}.
%
%%%%%%%%%%%%%%%%%%%%%%%%%%%%%%%%%%%%%%%%%%%%%%%%%%%%%%%%%%%%%%%%%%%%%%%%%%%%%%%%%%%%%%
%
\section{The Underdamped and Overdamped Limits}
\label{sec:asymptotics}
In this section we derive approximate formulas for the diffusion coefficient which are
valid in the overdamped $\gamma \gg 1 $ and underdampled $\gamma \ll 1$ limits. The
derivation of these formulas is based on the asymptotic analysis of the Poisson
equation~\eqref{eqn:cell}. In this section we will take the period of the potential is $2
\pi$.
\subsection{The Underdamped Limit}
In this subsection we solve the Poisson equation \eqref{eqn:cell} in one dimension
perturbatively for small $\gamma$. We shall use singular perturbation theory for partial
differential equations. The operator $\cL_0$ that appears in~\eqref{eqn:cell} can be
written in the form
$$
\cL_0 = \cL_H + \gamma \cL_{OU}
$$
where $\cL_H$ stands for the (backward) Liouville operator associated with the
Hamiltonian $H(p,q)$ and $\cL_{OU}$ for the generator of the OU process, respectively:
$$
\cL_H = p \partial_q - \partial_q V \partial_p, \quad \cL_{OU} = -p \partial_p +
\beta^{-1} \partial_p^2.
$$
We expect that the solution of the Poisson equation scales like $\gamma^{-1}$ when
$\gamma \ll 1$. Thus, we look for a solution of the form
\begin{equation}\label{e:ansatz}
\Phi = \frac{1}{\gamma} \phi_0 + \phi_1 + \gamma \phi_2 + \dots
\end{equation}
We substitute this ansatz in \eqref{eqn:cell} to obtain the sequence of equations
\begin{subequations}
\begin{eqnarray}
\cL_H \phi_0 &=& 0, \label{e:0} \\
-\cL_H \phi_1 & = & p +\cL_{OU} \phi_0 , \label{e:1} \\
-\cL_H \phi_2 & = & \cL_{OU} \phi_1. \label{e:2}
\end{eqnarray}
\end{subequations}
From equation \eqref{e:0} we deduce that, since the $\phi_0$ is in the null space of the
Liouville operator, the first term in the expansion is a function of the Hamiltonian
$z(p,q) =\frac{1}{2} p^2 + V(q) $:
$$
\phi_0 = \phi_0(z(p,q)).
$$
Now we want to obtain an equation for $\phi_0$ by using the solvability condition for
\eqref{e:1}. To this end, we multiply this equation by an arbitrary function of $z$, $g =
g(z)$ and integrate over $p$ and $q$ to obtain
$$
\int_{-\infty}^{+\infty} \int_{-\pi}^{\pi}  \left(p + \cL_{OU} \phi_0 \right) g(z(p,q))
\, dp dq = 0.
$$
We change now from $p,q$ coordinates to $z,q$, so that the above integral becomes
$$
\int_{E_{min}}^{+\infty} \int_{-\pi}^{\pi}  g(z)\left( p(z,q) + \cL_{OU} \phi_0(z)
\right) \frac{1}{p(z,q)} \, dz dq = 0,
$$
where $J = p^{-1}(z,q)$ is the Jacobian of the transformation. Operator $\cL_0$, when
applied to functions of the Hamiltonian, becomes:
$$
\cL_{OU} = (\beta^{-1} - p^2) \frac{\partial}{\partial z} + \beta^{-1} p^2
\frac{\partial^2}{\partial z^2}.
$$
Hence, the integral equation for $\phi_0(z)$ becomes
$$
\int_{E_{min}}^{+\infty} \int_{-\pi}^{\pi}  g(z)\left[ p(z,q) + \left( (\beta^{-1} - p^2)
\frac{\partial }{\partial z} + \beta^{-1} p^2 \frac{\partial^2}{\partial z^2} \right)
\phi_0(z) \right] \frac{1}{p(z,q)} \, dz dq = 0.
$$
Let $E_0$ denote the critical energy, i.e. the energy along the separatrix (homoclinic
orbit). We set
$$
S(z) = \int_{x_1(z)}^{x_2(z)} p(z,q) \, dq, \quad  T(z) = \int_{x_1(z)}^{x_2(z)}
\frac{1}{p(z,q)} \, dq,
$$
where  Risken's notation~\cite[p. 301]{Ris84} has been used for $x_1(z)$ and $x_2(z)$.

We need to consider the cases $\big\{ z > E_0, \, p> 0 \big\}, \; \big\{ z > E_0, \, p< 0
\big\}$ and $\big\{ E_{min} < z < E_0  \big\}$ separately.

We consider first the case $E > E_0, \, p> 0$. In this case $x_1(x)=\pi, \, x_2(z) =
-\pi$. We can perform the integration with respect to $q$ to obtain
$$
\int_{E_0}^{+ \infty} g(z)\left[2 \pi  + \left( (\beta^{-1} T(z) - S(z)) \frac{\partial
}{\partial z} + \beta^{-1} S(z) \frac{\partial^2}{\partial z^2} \right) \phi_0(z) \right]
\, dz = 0,
$$
This equation is valid for every test function $g(z)$, from which we obtain the following
differential equation for $\phi_0$:
\begin{equation}\label{e:phi_01}
-\overline{\cL} \phi:=-\beta^{-1} \frac{1}{T(z)} S(z) \phi'' + \left( \frac{1}{T(z)} S(z)
- \beta^{-1} \right) \phi' = \frac{2 \pi}{T(z)},
\end{equation}
where primes denote differentiation with respect to $z$ and where the subscript $0$ has
been dropped for notational simplicity.

A similar calculation shows that in the regions $E
> E_0, \, p< 0$ and $E_{min} < E < E_0$ the equation for $\phi_0$ is
\begin{equation}\label{e:phi_02}
- \overline{\cL} \phi = -  \frac{2 \pi}{T(z)}, \quad E > E_0, \, p< 0
\end{equation}
and
\begin{equation}\label{e:phi_03}
- \overline{\cL} \phi = 0, \quad E_{min} < E < E_0.
\end{equation}
Equations~\eqref{e:phi_01}, \eqref{e:phi_02}, \eqref{e:phi_03} are augmented with
condition~\eqref{e:l2_fin} and a continuity condition at the critical energy~\cite{FW99}
\begin{equation}\label{e:gluing}
2 \phi_3'(E_0) = \phi_1'(E_0) + \phi_2'(E_0),
\end{equation}
where $\phi_1, \, \phi_2, \, \phi_3$ are the solutions of equations~\eqref{e:phi_01},
\eqref{e:phi_02} and \eqref{e:phi_03}, respectively.

The average of a function $h(q,p) = h(q,p(z,q))$ can be written in the form~\cite[p.
303]{Ris84}
\begin{eqnarray*}
\langle h(q,p)\rangle_{\beta} &:=& \int_{-\infty}^{\infty} \int_{-\pi}^{\pi} h(q,p)
\mu_{\beta}(q,p) \, dq dp \nonumber \\ & = & Z_{\beta}^{-1} \int_{E_{min}}^{+\infty}
\int_{x_1(z)}^{x_2(z)} \Big(h(q,p(z,q)) + h(q,-p(z,q)) \Big) (p(q,z))^{-1} e^{-\beta z}
\, dz dq,
\end{eqnarray*}
where the partition function is
$$
Z_{\beta} = \sqrt{\frac{2 \pi}{\beta}} \int_{-\pi}^{\pi} e^{-\beta V(q)} \, dq.
$$
From equation~\eqref{e:phi_03} we deduce that $\phi_3 (z) =0$. Furthermore, we have that
$\phi_1(z) = -\phi_2(z)$. These facts, together with the above formula for the averaging
with respect to the Boltzmann distribution, yield:
\begin{eqnarray}
D & = & \langle p \Phi(p,q) \rangle_{\beta} = \langle p \phi_0 \rangle_{\beta} +
\mathcal{O}(1)
\\ & \approx  & \frac{2}{\gamma} Z_{\beta}^{-1} \int_{E_0}^{+ \infty} \phi_0(z) e^{\beta
z} \, dz \mathcal{O}(1) \nonumber \\ & = & \frac{4 \pi}{\gamma} Z_{\beta}^{-1}
\int_{E_0}^{+ \infty} \phi_0(z) e^{- \beta z} \, dz, \label{e:deff_energy}
\end{eqnarray}
to leading order in $\gamma$, and where $\phi_0(z)$ is the solution of the two point
boundary value problem~\eqref{e:phi_01}. We remark that if we start with formula $D =
\gamma \beta^{-1} \langle |\partial_p \Phi|^2  \rangle_{\beta}$ for the diffusion
coefficient, we obtain the following formula, which is equivalent
to~\eqref{e:deff_energy}:
$$
D = \frac{4 \pi}{\gamma \beta} Z_{\beta}^{-1} \int_{E_0}^{+ \infty} |\partial_z
\phi_0(z)|^2 e^{- \beta z} \, dz.
$$
Now we solve the equation for $\phi_0(z)$ (for notational simplicity, we will drop the
subscript $0$ ). Using the fact that $S'(z) = T(z)$, we rewrite~\eqref{e:phi_01} as
$$
- \beta^{-1} (S \phi')' + S \phi' = 2 \pi.
$$
This equation can be rewritten as
$$
- \beta^{-1} \big(e^{-\beta z} S \phi' \big) = e^{-\beta z}.
$$
Condition~\eqref{e:l2_fin} implies that the derivative of the unique solution
of~\eqref{e:phi_01} is
$$
\phi'(z) = S^{-1}(z).
$$
We use this in~\eqref{e:deff_energy}, together with an integration by parts, to obtain
the following formula for the diffusion coefficient:
\begin{equation}\label{e:deff_freidlin}
D = \frac{1}{\gamma} 8 \pi^2 Z_{\beta}^{-1} \beta^{-1} \int_{E_0}^{+\infty}
\frac{e^{-\beta z}}{S(z)} \, dz.
\end{equation}
We emphasize the fact that this formula is exact in the limit as $\gamma \rightarrow 0$
and is valid for all periodic potentials and for all values of the temperature.

Consider now the case of the nonlinear pendulum $V(q) = -\cos(q)$. The partition function
is
$$
Z_{\beta} = \frac{(2 \pi)^{3/2}}{\beta^{1/2}} J_0 (\beta),
$$
where $J_0(\cdot)$ is the modified Bessel function of the first kind. Furthermore, a
simple calculation yields
$$
S(z) = 2^{5/2} \sqrt{z +1} E \left( \sqrt{\frac{2}{z+1}} \right),
$$
where $E(\cdot)$ is the complete elliptic integral of the second kind. The formula for
the diffusion coefficient becomes
\begin{equation}\label{e:deff_cos}
D = \frac{1}{\gamma} \frac{\sqrt{\pi}}{2 \beta^{1/2} J_0(\beta)} \int_1^{+ \infty}
\frac{e^{- \beta z}}{\sqrt{z+1} E(\sqrt{2/(z+1)})} \, dz.
\end{equation}
We use now the asymptotic formula $J_0 (\beta) \approx (2 \pi \beta)^{-1/2} e^{\beta}, \;
\beta \gg 1$ and the fact that $E(1) = 1$ to obtain the small temperature asymptotics for
the diffusion coefficient:
\begin{equation}\label{e:deff_beta_large}
D = \frac{1}{\gamma} \frac{\pi}{2 \beta} e^{-2 \beta}, \quad \beta \gg 1,
\end{equation}
which is precisely formula~\eqref{e:risken}, obtained by Risken.

Unlike the overdamped limit which is treated in the next section, it is not
straightforward to obtain the next order correction in the formula for the effective
diffusivity. This is because, due to the discontinuity of the solution of the Poisson
equation~\eqref{eqn:cell} along the separatrix. In particular, the next order correction
to $\phi$ when $\gamma \ll 1$ is of $\mathcal(\gamma^{-1/2})$, rather than $\mathcal(1)$
as suggested by ansatz~\eqref{e:ansatz}.

Upon combining the formula for the diffusion coefficient and the formula for the hopping
rate from Kramers' theory~\cite[eqn. 4.48(a)]{HanTalkBork90} we can obtain a formula for
the mean square jump length at low friction. For the cosine potential, and for $\beta \gg
1$, this formula is
\begin{equation}\label{e:jump_length}
\langle \ell^2 \rangle = \frac{\pi^2}{8 \gamma^2 \beta^2} \quad \mbox{for} \; \; \gamma
\ll 1, \, \beta \gg 1.
\end{equation}

%
%%%%%%%%%%%%%%%%%%%%%%%%%%%%%%%%%%%%%%%%%%%%%%%%%%%%%%%%%%%%%%%%%%%%%%%%%%%%%%%%%%%%%%%
%
\subsection{The Overdamped Limit}
In this subsection we study the large $\gamma$ asymptotics of the diffusion coefficient.
As in the previous case, we use singular perturbation theory, e.g. \cite[Ch.
8]{HorsLef84}. The regularity of the solution of~\eqref{eqn:cell} when $\gamma \gg 1$
will enable us to obtain the first two terms in the $\frac{1}{\gamma}$ expansion without
any difficulty.

We set $\gamma = \frac{1}{\eps}$. The differential operator $\cL_0$ becomes
$$
\cL_0 = \frac{1}{\eps} \cL_{OU} + \cL_H.
$$
We look for a solution of~\eqref{eqn:cell} in the form of a power series expansion in
$\gamma$:
\begin{equation}\label{e:exp_overd}
\Phi = \phi_0 + \eps \phi_1 + \eps^2 \phi_2 + \eps^3 \phi_3 + \dots
\end{equation}
We substitute this into~\eqref{eqn:cell} and obtain the following sequence of equations:
\begin{subequations}\label{e:gamma_large}
\begin{eqnarray}
-\cL_{OU} \phi_0 & = & 0, \\
-\cL_{OU} \phi_1 & = & p + \cL_H \phi_0, \\
-\cL_{OU} \phi_2 & = & \cL_H \phi_1, \\
-\cL_{OU} \phi_3 & = & \cL_H \phi_2.
\end{eqnarray}
\end{subequations}
The null space of the Ornstein-Uhlenbeck operator $\cL_0$ consists of constants in $p$.
Consequently, from the first equation in~\eqref{e:gamma_large} we deduce that the first
term in the expansion in independent of $p$, $\phi_0 = \phi(q)$. The second equation
becomes
$$
-\cL_{OU} \phi_1 = p(1 + \partial_q \phi).
$$
Let
$$
\nu_\beta(p) = \left( \frac{2 \pi}{\beta}\right)^{-\frac{1}{2}} e^{- \beta
\frac{p^2}{2}},
$$
be the invariant distribution of the OU process (i.e. $\cL_{OU}^* \nu_{\beta}(p) = 0$).
The solvability condition for an equation of the form $-\cL_{OU} \phi = f$ requires that
the right hand side averages to $0$ with respect to $\nu_{\beta}(p)$, i.e. that the right
hand side of the equation is orthogonal to the null space of the adjoint of $\cL_{OU}$.
This condition is clearly satisfied for the equation for $\phi_1$. Thus, by Fredholm
alternative, this equation has a solution which is
$$
\phi_1(p,q) = (1 + \partial_q \phi) p + \psi_1(q),
$$
where the function $\psi_1(q)$ of is to be determined. We substitute this into the right
hand side of the third equation to obtain
$$
-\cL_{OU} \phi_2 = p^2 \partial_q^2 \phi - \partial_q V (1 + \partial_q \phi) + p
\partial_q \psi_1(q).
$$
From the solvability condition for this we obtain an equation for $\phi(q)$:
\begin{equation}\label{eqn:cell_smoluch}
\beta^{-1} \partial_q^2 \phi - \partial_q V (1 + \partial_q \phi) = 0,
\end{equation}
together with the periodic boundary conditions. The derivative of the solution of this
two-point boundary value problem is
\begin{equation}\label{e:soln_smoluch}
\partial_q \phi +1 = \frac{2 \pi}{\int_{-\pi}^{\pi} e^{\beta V(q)} \, dq } e^{\beta V(q)}.
\end{equation}
The first two terms in the large $\gamma$ expansion of the solution of
equation~\eqref{eqn:cell} are
$$
\Phi(p,q) = \phi(q) + \frac{1}{\gamma} (1 + \partial_q \phi) +
\mathcal{O}\left(\frac{1}{\gamma^2} \right),
$$
where $\phi(q)$ is the solution of~\eqref{eqn:cell_smoluch}. Substituting this in the
formula for the diffusion coefficient and using~\eqref{e:soln_smoluch} we obtain
\begin{eqnarray*}
D & = & \int_{-\infty}^{\infty} \int_{-\pi}^{\pi} p \Phi \rho_{\beta}(p,q) \, dp dq =
\frac{4 \pi^2}{\beta Z \widehat{Z}} + \mathcal{O}\left(\frac{1}{\gamma^3} \right),
\end{eqnarray*}
where $Z = \int_{-\pi}^{\pi} e^{- \beta V(q)}, \; \widehat{Z} = \int_{-\pi}^{\pi} e^{
\beta V(q)}$. This is, of course, the Lifson-Jackson formula which gives the diffusion
coefficient in the overdamped limit~\cite{lifson_jackson62}. Continuing in the same
fashion, we can also calculate the next two terms in the expansion~\eqref{e:exp_overd}.
From this, we can compute the next order correction to the diffusion coefficient. The
final result is
\begin{equation}
D = \frac{4 \pi^2 }{\beta \gamma Z \widehat{Z}} - \frac{4 \pi^2 \beta Z_1}{\gamma^3 Z
\widehat{Z}^2} + \mathcal{O} \left( \frac{1}{\gamma^5} \right), \label{e:deff_gamma}
\end{equation}
where $Z_1 = \int_{-\pi}^{\pi} |V'(q)|^2 e^{\beta V(q)} \, dq$.

In the case of the nonlinear pendulum, $V(q) = \cos(q)$, formula~\eqref{e:deff_gamma}
gives
\begin{equation}\label{e:large_gamma_cos}
D = \frac{1}{\gamma \beta} J_0^{-2} (\beta) - \frac{\beta}{\gamma^3} \left(
\frac{J_2(\beta)}{J_0^3(\beta)} - J_0^{-2}(\beta) \right) + \mathcal{O}
\left(\frac{1}{\gamma^5} \right),
\end{equation}
where $J_n(\beta)$ is the modified Bessel function of the first kind.

In the multidimensional case, a similar analysis leads to the large gamma asymptotics:
$$
\langle \xi, D \xi \rangle = \frac{1}{\gamma} \langle \xi, D_0 \xi \rangle + \mathcal{O}
\left( \frac{1}{\gamma^3} \right),
$$
where $\xi$ is an arbitrary unit vector in $\R^d$ and $D_0$ is the diffusion coefficient
for the Smoluchowski (overdamped) dynamics:
$$
D_0 = Z^{-1} \int_{\R^d} \big(- \cL_V \chi \big) \otimes \chi e^{- \beta V(q)} \, dq
$$
where
$$
\cL_V = - \nabla_q V \cdot \nabla_q + \beta^{-1} \Delta_q
$$
and $\chi(q)$ is the solution of the PDE $ \cL_V \chi = \nabla_q V$ with periodic
boundary conditions.

\section{The Numerical Method}
\label{sec:numerical}
In order to calculate the diffusion coefficient $D$ we have to solve
equation~\eqref{eqn:cell} and calculate the integral in~\eqref{eqn:eff_diff}. It is
possible to do this by means of a spectral method. In particular, by expanding the
solution of~\eqref{eqn:cell} into Hermite polynomials in $p$ and a standard Fourier
series in $q$, we can convert the boundary value problem into an infinite system of
linear equations; upon truncating this system we can obtain a finite dimensional, sparse
system of linear equations which we can easily solve. For simplicity we will consider the
problem in one dimension, though our numerical method works in arbitrary dimensions.

We look for a solution of~\eqref{eqn:cell} in the form
\begin{equation}\label{e:expansion}
\phi(p,q) = \sum_{k = -\infty}^{+ \infty}\sum_{n=0}^{+\infty} \phi_{n k} e^{i k q} f_n
(p),
\end{equation}
where $f_n$ denotes the $n$th eigenfunction of the Ornstein-Uhlenbeck operator
$\LL_{OU}$; the corresponding eigenvalue is $k$. The eigenfunctions of the OU processes
are related to the Hermite polynomials through the formula~\cite[Ch. 5]{Ris84}
\begin{equation}
f_n(p) = \frac{1}{\sqrt{ n!}} H_n \left( \sqrt{\beta} p \right), \label{e:hermite}
\end{equation}
where
$$
H_n(p) = (-1)^n e^{\frac{p^2}{2}} \frac{d^n}{d p^n} \left( e^{-\frac{p^2}{2}} \right).
$$
We remark that the solution of the Poisson equation~\eqref{eqn:cell} is defined up to a
constant which we have taken to be $0$ (i.e. we assume that the solution  averages to
zero with respect to the Maxwell-Boltzmann distribution). It is elementary to check that
the choice of this constant does not effect the value of the diffusion coefficient.
Notice also that the boundary conditions (the solution is periodic in $q$ and square
integrable with respect to the Maxwell-Boltzmann distribution) have already been taken
into account when writing~\eqref{e:expansion}.

Using the fact that the $p = \beta^{-1/2}f_1$, together with the orthonormality of the
eigenfunctions of the Ornstein-Uhlenbeck operator, we obtain the following formula for
the diffusion coefficient:
\begin{eqnarray}\label{e:deff_hermite}
D = Z^{-1} \beta^{-1/2} \sum_{k=-\infty}^{ \infty} \phi_{1 k} \int_{-\pi}^{\pi} e^{ik q}
e^{-\beta V(q)} \, dq.
\end{eqnarray}
For the cosine potential $V(q) =-\cos(q)$ the above formula becomes
\begin{equation}\label{e:deff_bessel}
D = \beta^{-1/2} J_0^{-1}(\beta) \sum_{k = -\infty}^{+ \infty} \phi_{1 k} J_k(\beta),
\end{equation}
Now we shall obtain a linear system of equations for the coefficients $\phi_{n k}$ for
the cosine potential. The Poisson equation~\eqref{eqn:cell} in one dimension is
\begin{equation}
- \LL \phi = p = \beta^{-1/2} f_1. \label{eqn:cell_2}
\end{equation}
We substitute the expansion~\eqref{e:expansion} into~\eqref{eqn:cell_2} and use the
properties of Hermite polynomials and of trigonometric functions to obtain the following
linear system of equations
\begin{subequations}
\label{e:lin_syst}
\begin{equation}
2 \beta^{-1} k \phi_{1 \, k} + \phi_{1 \, k-1} - \phi_{1 \, k+1} = 0,
\end{equation}
\begin{equation}
2 \gamma i \phi_{1 \, k}  + 2 \sqrt{\beta^{-1}} k \phi_{0 \, k} +  2 \sqrt{ 2\beta^{-1}}
k \phi_{2 \, k} + \sqrt{2 \beta} \phi_{2 \, k-1} - \sqrt{2 \beta} \phi_{2 \, k +1} = 2 i
\sqrt{\beta^{-1}} \, \delta_{k \, 0},
\end{equation}
\begin{eqnarray}
&&2 \sqrt{\beta^{-1}} \sqrt{n+1}  k \phi_{n+1 \, k} + 2 \sqrt{\beta^{-1}} \sqrt{n}  k
\phi_{n-1 \, k} \nonumber \\ &&+ \sqrt{\beta} \sqrt{n+1} \phi_{n+1 \, k-1} - \sqrt{\beta}
\sqrt{n+1} \phi_{n+1 \, k+1} +2 i \gamma n \phi_{n \, k} = 0.
\end{eqnarray}
\end{subequations}
\begin{figure}\label{fig:deff}
\begin{center}
\includegraphics[width=3.2in, height = 3.2in]{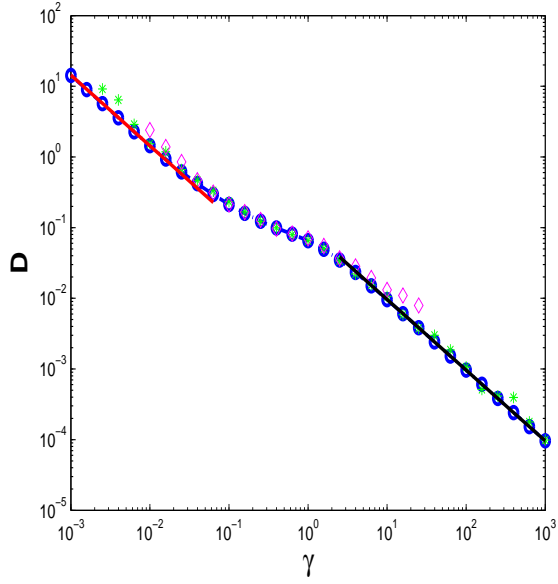}
\caption{Diffusion coefficient as a function of the friction coefficient for the cosine
potential. Dash-dot line and circles: $D$ obtained  from the numerical solution of the
Poisson equation, formula~\eqref{e:deff_bessel}; stars: $D$ obtained from the the
calculation of the jump length distribution and the hopping rate,
formula~\eqref{e:deff_jump};diamonds: results from Monte Carlo simulations,
formula~\eqref{e:deff_langr};solid lines, analytical approximation for $\gamma \ll 1, \;
\gamma \gg 1$, equations~\eqref{e:deff_beta_large} and~\eqref{e:lif_jack_beta}. }
\end{center}
\end{figure}
We truncate~\eqref{e:lin_syst} by taking into account the first $N + 1$ terms of the
Hermite expansion and the first $K + 1$ terms of the Fourier expansion. This leads to a
sparse linear system of $\brk{N+1}\brk{2K+1}$ equations and $\brk{N+1}\brk{2K+1}-1$
variables (note the absence of the term $\fnk{0}{0}$). This implies that one of the
equations in~\eqref{e:lin_syst} is linearly dependent on the others; to obtain a
nonsingular system we remove the equation for $k = 0$.\footnote{This follows from our
assumption that the solution of the Poisson equation averages to $0$ with respect to the
canonical ensemble.} In this way we obtain a nonsingular system of
$s=\brk{N+1}\brk{2K+1}-1$ equations and unknowns. This system can be written in the form
$\mathbf{Ax=b}$ with
\begin{equation}\nonumber
x_i=\fnk{n}{k},\hspace{0.2cm} i=\left\{
                    \begin{array}{ll}
                      k + K + 1 & n=0, k < 0, \\
                      n\brk{2K+1}+k+K & n=0, k > 0$ or $n\neq0\\
                    \end{array}
                  \right .
\end{equation}
and $\alpha_{i\,j}$ is the coefficient of $x_j$ in the $i^{th}$ equation, which is taken
from equations~\eqref{e:lin_syst} for values of $n$ and $k$
$$
i=\left\{
                    \begin{array}{ll}
                      k + K + 1 & n=0, k < 0 \\
                      n\brk{2K+1}+k+K & n=0, k > 0$ or $n\neq0\\
                    \end{array}
                  \right .
$$
The sparsity of this system implies that we can solve it very efficiently. In particular,
we can calculate accurately the diffusion coefficient with a minimal computational cost,
even for very small values of the friction constant $\gamma$.

To illustrate the efficiency of our numerical method, we calculate the diffusion
coefficient for the cosine potential as a function of the dissipation $\gamma$, at a
fixed temperature $\beta^{-1} = 0.5$. We compare the results obtained through our
numerical method with the approximate analytical expressions~\eqref{e:lif_jack_beta} and
$~\eqref{e:risken}$, results from Monte Calro simulations using~\eqref{e:deff_langr} and
results obtained through numerical calculation of the hopping rate and the jump length
distribution, equation~\eqref{e:deff_jump}. FOr the calculation of $\langle \ell^2
\rangle$ and $\kappa$ we generate a long path (for every $\gamma$) of the Langevin
dynamics using the Milstein scheme.

The results of the numerical simulations are presented in Figure~\ref{fig:deff}. There
agreement between the approximate analytical formulas and the calculation of the
diffusion coefficient using the method described in this section are excellent. Our
method is far superior in comparison to Monte Carlo simulations or the calculation of the
mean square jump length and the hopping rate, since even for very small $\gamma$ the
solution of a rather small linear system of equations is required.\footnote{Needless to
say, more Hermite and Fourier terms have to be taken into account when $\gamma$
decreases. However, even for $\gamma$ very small, the resulting linear system of
equations is small enough so that it can be solved in a few seconds in Matlab.} On the
other hand, contrast, the path of integration over which we have to solve the Langevin
equation in order to compute accurate statistics increases as $\gamma$ decreases and the
calculation of $D$ using Monte Carlo becomes computationally expensive.
%%%%%%%%%%%%%%%%%%%%%%%%%%%%%%%%%%%%%%%%%%%%%%%%%%%%%%%%%%%%%%%%%%%%%%%%%%%%%%%%%%%%%
%
\section{Conclusions}
\label{sec:conclusions} The problem of Brownian motion in a periodic potential in
arbitrary dimensions was studied in this paper. Using multiscale
techniques~\cite{PavlSt08} we derived a formula for the effective diffusion tensor which
is valid for all values of the friction coefficient and the temperature, and in arbitrary
dimensions. We also showed the equivalence between our formula and the Green-Kubo
formula. The calculation of the diffusion tensor using our approach requires the solution
of a Poisson equation together with the calculation of the average of an appropriate
function with respect to the canonical distribution. Furthermore, the overdamped and
underdamped limits where studied and approximate analytical formulas for the diffusion
coefficient in these two limits where obtained. In addition, a very efficient numerical
method for the calculation of the diffusion coefficient was developed; this numerical
method is based on the solution of the Poisson equation via a spectral method and it
leads to the accurate and very efficient calculation of the diffusion coefficient even
for very low values of the friction coefficient.

The approach developed in the paper for the study of the problem of diffusion in periodic
potentials offers various advantages over other analytical and numerical methods. First,
all the results reported in this paper can be justified rigorously and they can also lead
to a rigorous analysis of the dependence of the diffusion tensor on the friction
coefficient and on the temperature. A first step in this direction was taken
in~\cite{HP07}. Second, our method enables us to study various distinguished limits of
physical interest (such as the overdamped and underdamped limits) in a systematic fashion
through asymptotic analysis of the Poisson equation. Third, it leads to an efficient
numerical method for calculating the diffusion tensor through the numerical solution of
the Poisson equation. The effectiveness of our method was shown in this paper for the one
dimensional problem, for which analytical approximate formulas are can be derived. The
method is also very efficient in two and three dimensions and can offer insight into the
problem of Brownian motion in periodic potentials in higher dimensions. A thorough
numerical investigation of the multidimensional problem will be presented elsewhere.

\section*{Acknowledgements}
The authors are grateful to Igor Goychuk for useful suggestions and comments.

\def\cprime{$'$} \def\cprime{$'$} \def\cprime{$'$} \def\cprime{$'$}
  \def\cprime{$'$} \def\cprime{$'$} \def\cprime{$'$}
  \def\Rom#1{\uppercase\expandafter{\romannumeral #1}}\def\u#1{{\accent"15
  #1}}\def\Rom#1{\uppercase\expandafter{\romannumeral #1}}\def\u#1{{\accent"15
  #1}}\def\cprime{$'$} \def\cprime{$'$} \def\cprime{$'$} \def\cprime{$'$}
  \def\cprime{$'$} \def\cprime{$'$} \def\cprime{$'$}

\end{document}